\documentclass[prb,aps,reprint,floatfix]{revtex4-1}
\usepackage[utf8]{inputenc}
\usepackage{graphicx}
\usepackage{subfigure}
\usepackage{dcolumn}
\usepackage{amsmath}
\usepackage{gensymb}
\usepackage{textcomp}

\begin{document}

\title{Substrate-induced suppression of charge density wave in monolayer 1H-TaS$_2$ on Au(111)}
\author{Heraclitos M. Lefcochilos-Fogelquist}
\affiliation{
    Department of Physics,
    Georgetown University,
    Washington, DC 20057, USA
}
\author{Oliver R. Albertini}
\affiliation{
    Department of Physics,
    Georgetown University,
    Washington, DC 20057, USA
}
\author{Amy Y. Liu}
\affiliation{
    Department of Physics,
    Georgetown University,
    Washington, DC 20057, USA
}

\date{\today}

\begin{abstract}
Recent experiments have found that monolayer 1H-TaS$_2$ grown on Au(111) lacks the charge density wave (CDW) instability exhibited by bulk 2H-TaS$_2$. Additionally, angle-resolved photoemission spectroscopy measurements suggest that the monolayer becomes strongly electron doped by the substrate. While density functional theory (DFT) calculations have shown that electron doping can suppress the CDW instability in monolayer 1H-TaS$_2$,  it has been suggested that the actual charge transfer from the substrate may be much smaller than the apparent doping deduced from photoemission data. We present DFT calculations of monolayer 1H-TaS$_2$ on Au(111) to explore  substrate effects beyond doping. We find that the CDW instability is suppressed primarily by strong S-Au interactions rather than by doping. The S-Au interaction results in a structural distortion of the TaS$_2$ monolayer characterized by both lateral and out-of-plane atomic displacements and a $7 \times 7$ periodicity dictated by the commensurate interface with Au. Simulated STM images of this $7\times 7$ distorted structure are consistent with experimental STM images. In contrast, we find a robust $3 \times 3$ CDW phase in monolayer 1H-TaS$_2$ on a graphene substrate with which there is minimal interaction.
\end{abstract}
\maketitle

\section{\label{sec::intro}Introduction}

Transition metal dichalcogenides (TMDCs) form a family of layered materials with accessible single-layer (SL) phases, many of which host interesting physical properties and have potential applications in nanoscale electronics. A number of metallic TMDCs  exhibit symmetry-breaking electronic instabilities such as charge-density-wave (CDW) and superconducting transitions,\cite{wilson,Wilson-Yoffe-TMDCreview} and the evolution of these instabilities in the two-dimensional (2D) limit is an active area of investigation. In the case of SL NbSe$_2$, for example, the stability and structure of the CDW phase remain under debate, as first-principles calculations\cite{calandra-mazin-mauri} and different  experiments\cite{mak-NbSe2,crommie-NbSe2} have yielded conflicting results. Discrepancies like this could arise from differences in the substrate used (and the lack of a substrate in the calculations), as 2D materials can be highly sensitive to their environment. Indeed, it has been shown that effects such as oxidation can suppress CDW formation in other SL and few-layer TMDCs such as 1T-TaS$_2$.\cite{tsen,Wang-1TTaS2CDW,albertini-1TTaS2phonons} Therefore, in probing the behavior of 2D TMDCs it is crucial to distinguish between intrinsic properties and effects induced by the ambient environment or substrate.

\par  

To avoid oxidation issues, SL TaS$_2$ was recently grown epitaxially on Au under ultra-high vacuum conditions.\cite{sanders,Wehling} This approach had previously been used to grow large-area and high-quality SL crystals of the semiconducting TMDCs MoS$_2$ and WS$_2$.\cite{Gronborg-MoS2-Langmuir,Dendzik-WS2} The Au support was found to have a weak impact on the atomic structure of the semiconducting TMDCs, though evidence for hybridization between S and Au orbitals was observed in the valence bands.\cite{Sorensen,Bruix-MoS2-Au-PRB} In the case of TaS$_2$ on Au(111), it was found that the monolayer grows in the H polytype.
In contrast to bulk 2H-TaS$_2$, which adopts a $3\times3$ CDW structure characterized primarily by in-plane displacements of Ta atoms below about 75 K,\cite{Wilson-Yoffe-TMDCreview} scanning tunneling microscopy (STM) and low-energy electron diffraction  data show no evidence of a CDW transition in SL TaS$_2$ on Au, at least down to 5 K. \cite{sanders} Additionally, angle-resolved photoemission spectroscopy (ARPES) data suggests that the monolayer becomes n-doped by about 0.3 electrons per unit cell by the Au substrate.\cite{sanders}
Conversely, other experimental work has reported a robust $3\times 3$ CDW phase in SL 1H-TaS$_2$ grown by molecular beam epitaxy on bi-layer graphene.\cite{Lin-MBE-TMDonGraphene}
These experimental results show that the choice of substrate can affect CDW formation in SL 1H-TaS$_2$.

Density functional theory (DFT) calculations have found a freestanding SL of 1H-TaS$_2$ to be unstable to a $3 \times3$ lattice distortion similar to the bulk,\cite{Albertini-TaS2doping,Duxbury-TaX2}  indicating that the observed suppression of the CDW on Au is likely induced by the substrate. Calculations also found that the instability in a freestanding SL can be progressively removed with electron doping, supporting the suggestion that charge transfer from the metallic support can suppress the CDW instability.\cite{Albertini-TaS2doping} On the other hand, a recent study that combines DFT calculations with scanning tunneling spectroscopy to probe electronic states above the Fermi level found that hybridization between 1H-TaS$_2$ and Au valence states can induce considerable changes to the Fermi surface of the monolayer, leading to an apparent doping level that is significantly larger than the actual charge transfer.\cite{Wehling} This raises into question the role of doping in the observed suppression of the CDW.

\par
In this paper we use DFT calculations to investigate how the presence of a Au(111) substrate affects CDW formation in SL 1H-TaS$_2$ beyond simple doping effects. We find significant interaction between the TaS$_2$ monolayer and Au substrate, resulting in both lateral and out-of-plane displacements of atoms in the TaS$_2$ monolayer. The periodicity of the distorted structure is dictated by the $7 \times 7$ commensurate interface with Au rather than by the ordering vector for the $3\times 3$ CDW phase of the freestanding monolayer. Simulated STM images and their Fourier transforms are consistent with experiment when this structural distortion is included. For comparison, calculations for 1H-TaS$_2$ on a graphene substrate are also presented, where we find the $3\times 3$ CDW phase to be robust and the monolayer-substrate interaction to be weak.

\section{\label{sec:methods}Methods}

Density-functional theory calculations were carried out using the VASP package.\cite{vasp1} In most of the calculations, the exchange-correlation interaction was treated using the Perdew-Zunger local density approximation (LDA),\cite{perdew-zunger} but tests were also done using the OPTB88-vdW van der Waals density functional (vdW-DF).\cite{OPTB88-vdW}  The projected augmented wave method \cite{PAW-1994,PAW-1999} was used in conjunction with plane wave basis sets with cutoff energies of 260 eV (LDA) and 400 eV (OPTB88-vdW). Atomic optimization calculations were conducted using a minimum force threshold of $10^{-2}$ eV/\r{A}.

\par 
Experimentally, 1H-TaS$_2$ has been found to grow epitaxially on Au(111) with a $0\degree$ rotation between the overlayer and the substrate and a coincidence ratio of 7 TaS$_2$ cells to 8 Au(111) cells, corresponding to a coincidence mismatch of -0.6\%.\cite{sanders}  In addition to this experimentally observed interface, which we will refer to as the $7\times 7$ supercell, we have also considered an interface with a $30\degree$ rotation, where a $3\times3$ TaS$_2$ cell aligns with a $2\sqrt3\times2\sqrt3$ Au(111) cell with a coincidence lattice mismatch of 0.4\%. Although this rotated interface does not correspond to the monolayer-substrate orientation seen in experiments, it is a useful supercell to consider since its periodicity admits the CDW phase of the freestanding monolayer. Note that in the absence of a symmetry-lowering lattice distortion, the rotated interface can equivalently be described by a smaller  $\sqrt3\times\sqrt3$ TaS$_2$ supercell on a $2\times2$ Au(111) supercell.

\par
The Au(111) surface was modeled using a slab of five atomic layers terminated with H atoms on the side opposite the TaS$_2$ monolayer. 
At least 12 \r{A}  of vacuum were used to separate periodic images of the system. The in-plane lattice constant of 1H-TaS$_2$ was fixed to the experimental value of 3.316 \r{A}. The positions of atoms in the TaS$_2$ monolayer were allowed to fully relax, and Au atoms at the interface were allowed to relax normal to the surface (in the $z$ direction). The positions of other atoms in the Au slab were fixed. 

To assess the adequacy of the number of Au layers used, we tested the $\sqrt3\times\sqrt3$ supercell using Au slabs of up to 32 atomic layers.  The relaxed structure is nearly identical whether the Au surface is modeled with five layers or 32 layers.  Regarding the electronic structure,  periodic boundary conditions in the out-of-plane direction lead to the appearance of $k_z$ subbands arising from bulk Au conduction bands. This introduces spurious avoided crossings in the band structure. These artifacts are easily identified by using the system with 32 Au layers as a reference.\cite{supplemental} The spurious avoided crossings are found to have negligible effect on the simulated STM results.

To explore the effects of a graphene substrate, calculations were conducted using a $3 \times 3$ supercell of 1H-TaS$_2$ on a $4 \times 4$ supercell of graphene. As was the case for the Au(111) substrate, the in-plane lattice constant of 1H-TaS$_2$ was fixed to 3.316 \r{A}, leading to a strain of 0.28\% in the graphene layer. All atomic positions were optimized.

\section{\label{sec:results}Results \& Discussion}

\subsection{\label{sec:structure}Atomic structure of TaS$_2$ on Au}

\begin{figure}[tb]
  \includegraphics[width=0.85\linewidth,clip]{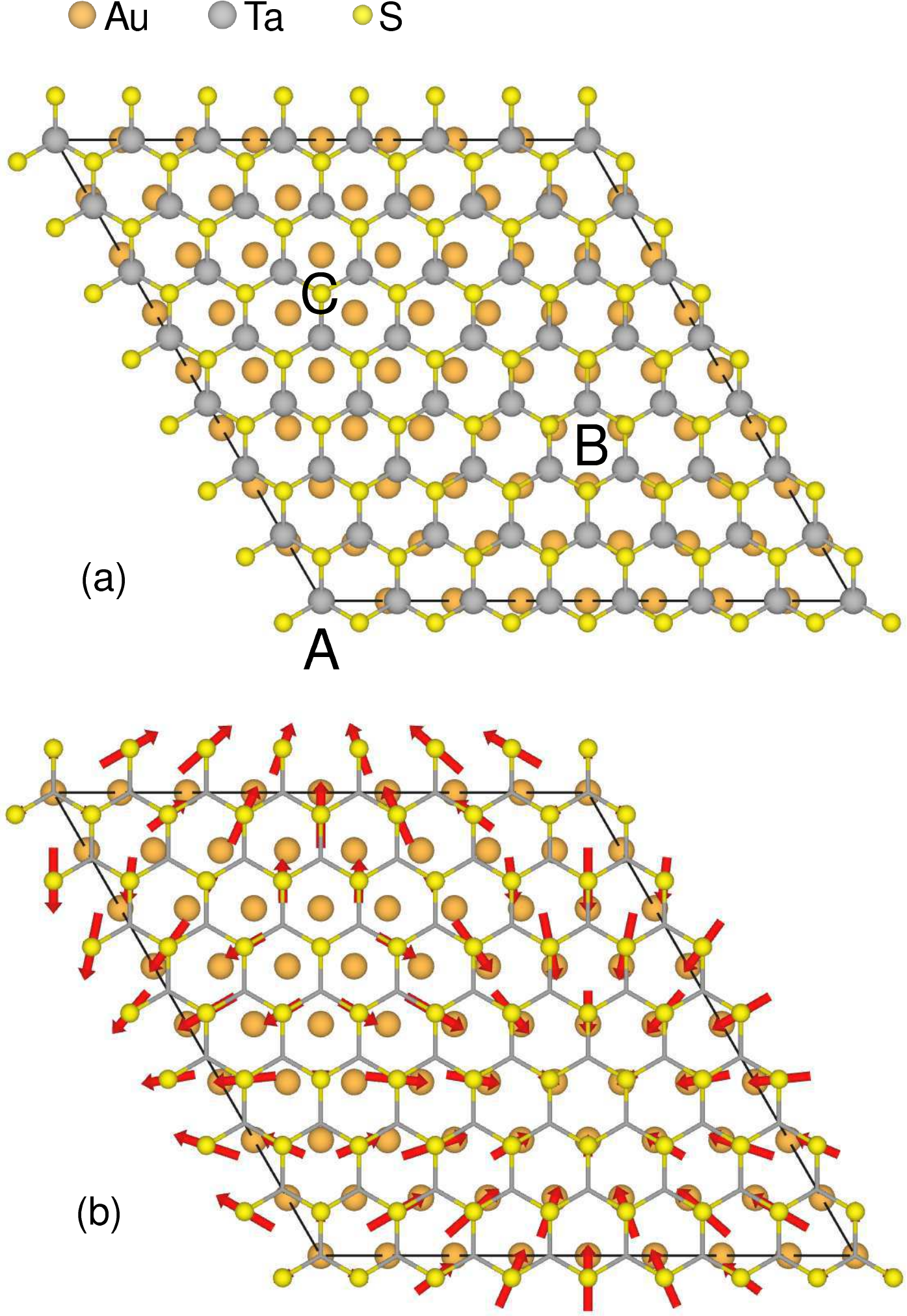}
  \caption{Top view of $7\times7$ supercell, showing TaS$_2$ monolayer and surface Au layer. (a) In region A, Ta atoms are approximately above Au surface sites, in region B, S atoms are approximately above Au surface sites, and in region C, both S and Ta atoms are approximately above surface hollow sites. (b) Arrows indicate lateral displacements of interface S atoms (not to scale). Ta atoms are omitted for clarity.}
  \label{fig:struct}
\end{figure}

The 1H-TaS$_2$ monolayer and the top layer of Au atoms at the interface in the $7\times7$ supercell are shown in Fig. \ref{fig:struct}(a).  Near the corner of the supercell, labeled region A, Ta atoms are aligned nearly on top of surface Au sites, while S atoms are aligned with second-layer Au atoms (hcp hollow sites). In region B, the Ta atoms align with third layer Au atoms (fcc hollow sites), while S atoms are nearly on top of surface Au atoms. In region C, the Ta atoms are above hcp hollow sites and the S atoms are above fcc hollow sites. While Fig. \ref{fig:struct} depicts a lateral alignment in which one Ta site is directly above a surface Au site, the supercell is large enough that even with an arbitrary lateral shift, there would still be regions with monolayer-substrate  alignments similar to regions A, B, and C.

When the TaS$_2$ and surface Au atoms are allowed to relax in the $z$ direction only, calculations find an average vertical separation of 2.42 \r{A} between the overlayer and substrate. Nearest-neighbor S-Au distances range from 2.45 \r{A} in region B to 2.98 \r{A} in   region C.  The overlayer remains relatively flat, with the $z$ coordinates of top-layer S atoms varying by less than 0.06 \r{A}.

When the TaS$_2$ atomic positions are allowed to relax laterally as well as vertically, the average $z$ separation between the monolayer and substrate decreases slightly, but more importantly, the monolayer develops significant ripples.
The pattern of rippling correlates with the
lateral proximity of S and Au atoms at the interface.  Peaks in B regions, where interface S atoms are nearly on top of surface Au sites, are surrounded by troughs in A and C regions. This rippling, with a height difference of nearly 0.4 \r{A} between the B and A regions is accompanied by lateral displacements of up to 0.15 \r{A} within the interface S layer. In the Ta layer, lateral displacements of up to 0.06 \r{A} are found. While the Ta displacements are similar in magnitude to those  calculated for the CDW phase of freestanding SL 1H-TaS$_2$, the pattern of Ta displacements does not resemble the CDW pattern.

The red arrows in Fig. \ref{fig:struct}(b) show the direction and relative size of lateral displacements of the interface S atoms. It is clear that the interface S atoms laterally displace toward the closest surface Au atoms. The vertical and lateral displacements combine to narrow the distribution of S-Au distances at the interface, leading to S-Au nearest-neighbor distances ranging from 2.52 \r{A} in region B to 2.85 \r{A} in region C.
The calculated distortion pattern and the resulting interatomic distances at the interface suggest significant interaction between the substrate and the overlayer.

For comparison, we carried out calculations using the van der Waals corrected OPTB88-vdW functional to see how sensitive the structural distortions are to the type of functional used. The LDA tends to overbind, so it is not surprising that the vdW-DF yields a larger average vertical separation of 2.74 \r{A} between the substrate and monolayer. Yet the pattern of rippling and lateral displacements remains qualitatively very similar, though the magnitudes of the displacements are reduced to roughly half the LDA values.

\begin{figure}[t]
  \includegraphics[width=0.85\linewidth,clip]{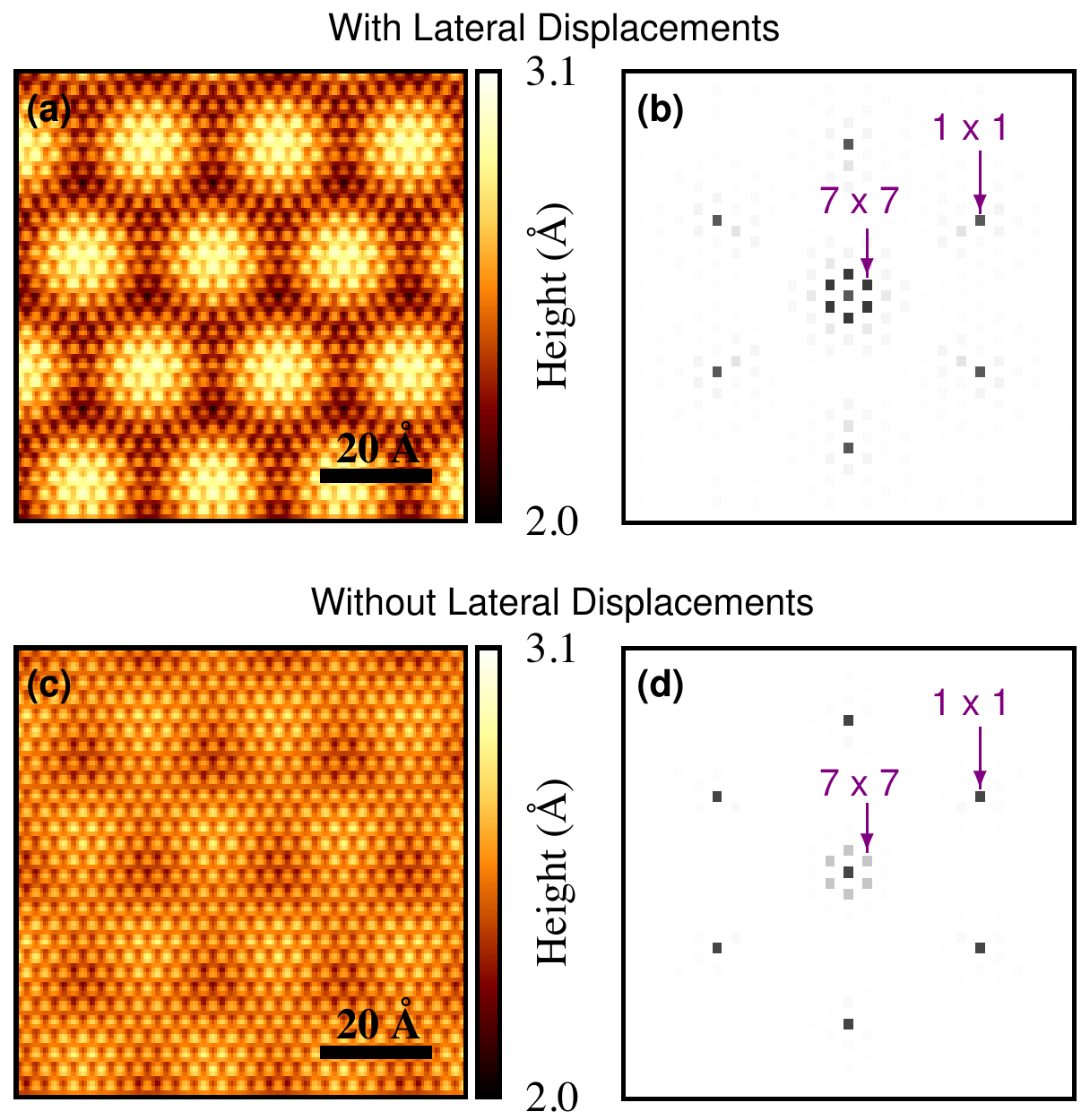}
  \caption{Simulated STM images. (a) Height map calculated from local density of states isosurface for fully relaxed structure. (b) FFT of image in (a). (c) Height map calculated from local density of states isosurface for structure without lateral displacements. (d) FFT of image in (c). The constant term represented by the point at the origin in the FFT images is set to the value of the 1x1 signal for clarity.}
  \label{fig:STM-DFT}
\end{figure}

Experimentally, the STM constant-current height map of monolayer TaS$_2$ on Au(111) shows a hexagonal superstructure with lattice constant 23.1 \r{A},  corresponding to the $7\times 7$ periodicity of the commensurate interface.\cite{sanders} The $7\times 7$ periodicity registers strongly in the Fast Fourier Transform (FFT) of the STM image --- more strongly, in fact, than the $1 \times 1$ TaS$_2$ periodicity. In experiment, this $7\times 7$ superstructure is attributed to a Moir\'{e} pattern arising from the lattice mismatch at the interface.

Figure \ref{fig:STM-DFT} shows our simulated STM images of both the fully relaxed and the laterally constrained structures. These images were rendered from isosurfaces corresponding to the same value of the local density of states (LDOS) for both structures.\cite{Tersoff-Hamann} The STM images of both structures exhibit a $7\times 7$ superstructure whose bright areas align with the B regions; however the pattern of this superstructure differs qualitatively depending on whether lateral atomic displacements are included. The pattern of the fully relaxed  structure qualitatively matches the honeycomb pattern seen in experiment,\cite{sanders} while that of the laterally constrained structure does not. In Fourier space the signal corresponding to the $7 \times 7$ superstructure registers clearly for both structures, but it is about three times stronger in the fully relaxed case than in the constrained case.

These differences between the fully relaxed and laterally constrained cases are robust in that they are relatively insensitive to the choice of LDOS isovalue. For a reasonable range of isosurfaces tested, the patterns of the $7\times 7$ superstructures in real space do not qualitatively change.  The existing $7\times 7$ patterns become more pronounced for smaller LDOS isovalues (corresponding to isosurfaces farther from the monolayer surface) since larger scale structure decays less rapidly in vacuum.\cite{STM-Kobayashi} In Fourier space this corresponds to an increase in the $7\times 7$ signal strength relative to the $1\times 1$ signal. For the full range of reasonable isovalues tested, the FFT of the STM image for the fully relaxed structure is compatible with that of the experiment, in which the $7\times 7$ signal is stronger than the $1\times 1$ signal and convolution spots appear around the $1\times 1$ signal. For the laterally constrained structure, there is a small range of isovalues for which the $7 \times 7$ signal is stronger than the $1\times 1$ signal, but the corresponding real-space STM image remains incompatible with the experimental results.  Hence the STM data, combined with our calculations, provide support for a structural distortion of the TaS$_2$ monolayer beyond that typically associated with a Moir\'{e} pattern.

\par
Energetically, we find that this distortion in the atomic structure is strongly favorable. The presence of lateral distortions provides an energy gain of 19 meV/Ta, which is roughly an order of magnitude greater than the energy gain calculated for the $3 \times 3$ CDW distortion in freestanding 1H-TaS$_2$. Based on the calculations presented thus far, however, we are not able to determine whether the CDW structure is dynamically unstable on the Au substrate or just energetically metastable, since the $7\times 7$ supercell does not accommodate the CDW periodicity.

\begin{figure}[tb]
  \includegraphics[width=0.85\linewidth,clip]{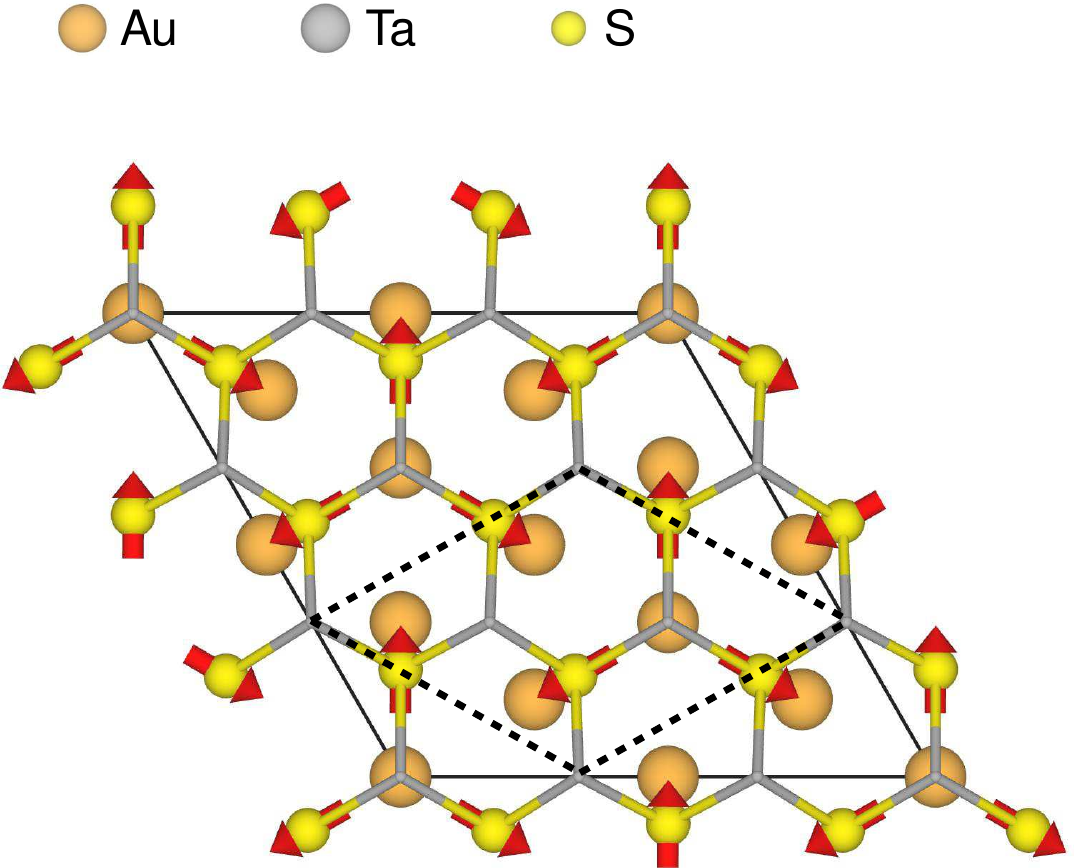}
  \caption{Top view of $3\times 3$ supercell, showing TaS$_2$ monolayer and surface Au layer. Arrows indicate lateral displacements of interface S atoms (not to scale). The optimized structure can also be described by a $\sqrt{3}\times\sqrt{3}$ TaS$_2$ on $2\times 2$ Au(111) supercell indicated with dashed lines. Ta atoms are omitted for clarity.}
  \label{fig:struct3x3}
\end{figure}

\par
To address the dynamical stability of the CDW structure of 1H-TaS$_2$ on Au, a supercell consisting of $3\times 3$ TaS$_2$ on $2\sqrt{3}\times2\sqrt{3}$ Au(111) was considered. Calculations were run with different initial structures, including undistorted 1H-TaS$_2$ atomic positions, the $3\times 3$ CDW structure calculated for freestanding 1H-TaS$_2$, and structures with different lateral alignment between the substrate and monolayer. Regardless of the initial configuration, the system relaxed to a structure that does not resemble the freestanding $3\times 3$ CDW phase. Instead, as shown in Fig. \ref{fig:struct3x3}, we find substantial lateral displacements of interface S atoms towards the closest Au atom (up to 0.08 \r{A}), qualitatively similar to what we found in the $7 \times 7$ supercell. This is notable as the $3 \times 3$ supercell has a different monolayer-substrate orientation than what is found in experiments. The similarities in the displacement of interface S atoms relative to the interface Au atoms strongly suggests that the S-Au interaction drives the structural distortion in both cases and causes the CDW structure to become dynamically unstable.

\begin{figure}[b]
  \includegraphics[width=0.9\linewidth,clip]{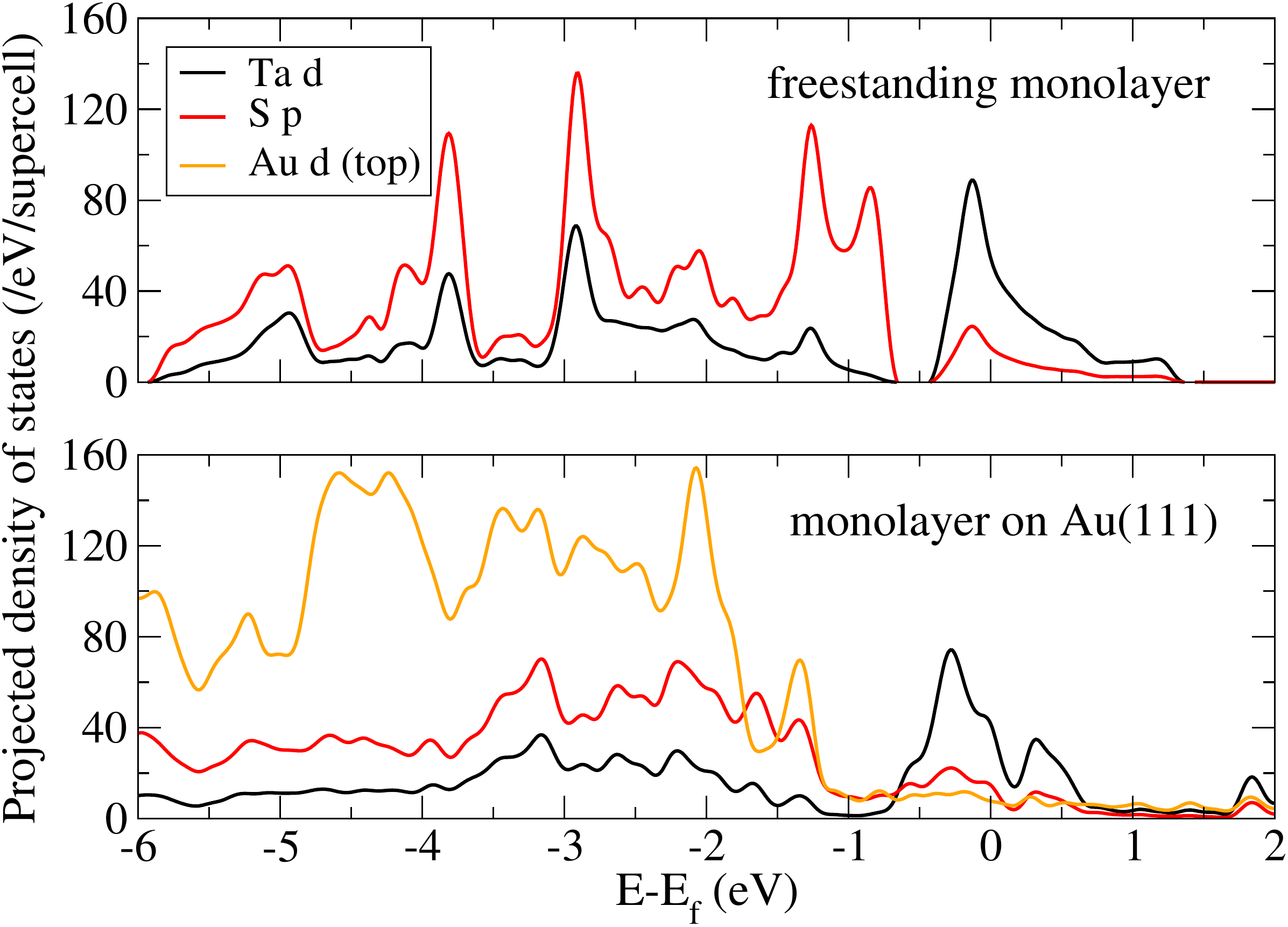}
  \caption{Projected density of states calculated for freestanding undistorted 
  1H-TaS$_2$ (top) and relaxed 7x7 supercell of 1H-TaS$_2$ on Au(111) (bottom).}
  \label{fig:pdos}
\end{figure}

\subsection{Electronic structure of TaS$_2$ on Au}

\begin{figure}[t]
  \includegraphics[width=0.9\linewidth,clip]{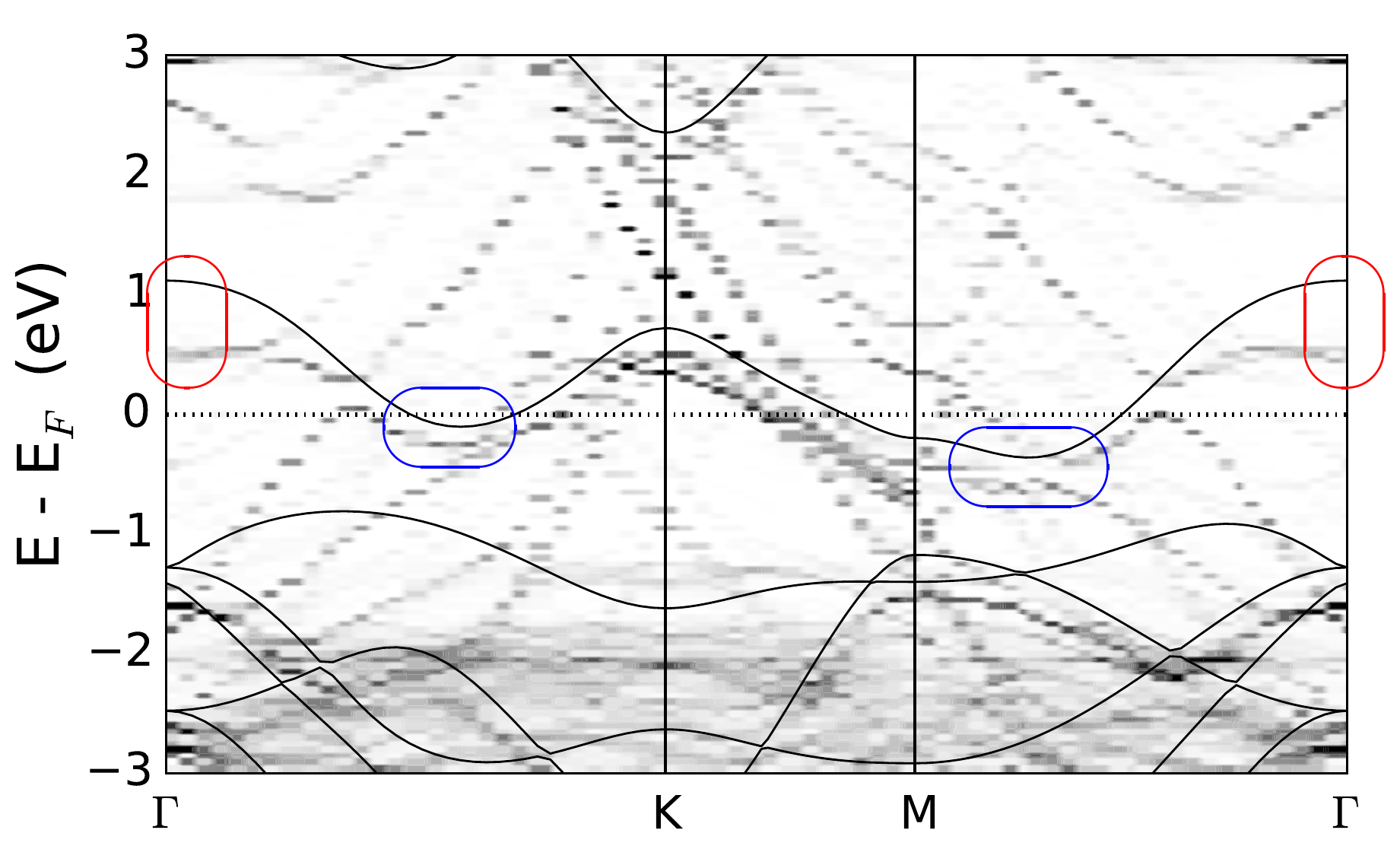}
  \caption{Band structure of $7\times 7$ 1H-TaS$_2$ on Au(111) supercell unfolded into $1\times 1$ TaS$_2$ Brillouin zone. Bands of free-standing 1H-TaS$_2$ monolayer are plotted with solid lines.  } 
  \label{fig:band-unfold}
\end{figure}

The orbital projected DOS (pDOS) of the undistorted freestanding monolayer and the relaxed $7\times7$ supercell of TaS$_2$ on Au are shown Fig. \ref{fig:pdos}. 
In the freestanding monolayer, an isolated band crosses the Fermi level, separated from a manifold of six filled bands that lie in the range of -6 eV to -1 eV (solid lines in Fig. \ref{fig:band-unfold}). The pDOS shows that the band at the Fermi level is dominated by Ta $d$ character, but has some S $p$ weight, while the manifold of bands below the Fermi level are primarily S $p$ in nature, with some Ta weight.

In the presence of the Au(111) substrate, the monolayer pDOS changes significantly. In particular, the shape of the S $p$ pDOS below the Fermi level is strongly influenced by the Au $d$ states, indicating strong S-Au hybridization. The bottom of the Ta $d$ band is shifted downward by roughly 0.18 eV, and the occupied region of the shifted Ta $d$ pDOS remains similar in  shape to that of the freestanding band. 
Above the Fermi level, there are two main changes in the Ta pDOS: a valley develops just above the Fermi level and the top of the band is shifted downward by about 0.5 eV. The former is an artifact of using a slab geometry to represent the interface (as discussed in Sec. \ref{sec:methods}), but the latter reflects a real change in the band structure.\cite{supplemental}

\begin{figure}[t]
  \includegraphics[width=0.85\linewidth,clip]{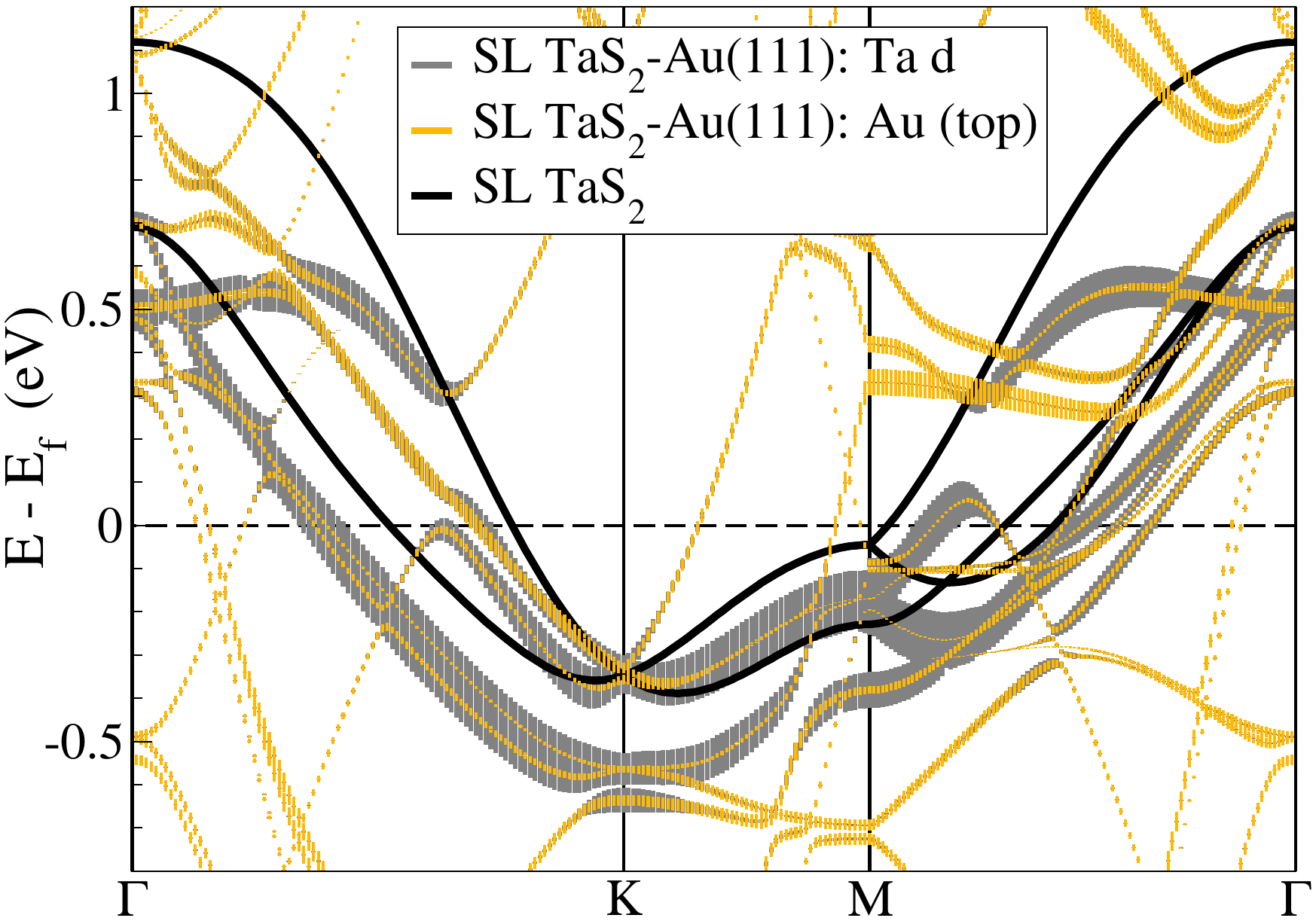}
  \caption{Orbital projected band structure of $\sqrt{3}\times\sqrt{3}$ TaS$_2$ on Au(111) supercell. Bands of free-standing 1H-TaS$_2$ monolayer are plotted with solid lines. }
  \label{fig:Au-PBands}
\end{figure}

To further explore substrate-induced changes in the electronic structure, we show the unfolded band structure\cite{BandUp1,BandUp2} of the $7\times7$ supercell in Fig. \ref{fig:band-unfold} and the orbital projected band structure of the $\sqrt{3}\times\sqrt{3}$ supercell in Fig. \ref{fig:Au-PBands}. As discussed in Sec. \ref{sec:structure}, the $\sqrt{3}\times\sqrt{3}$ supercell undergoes an atomic distortion which appears to be driven by the same interaction with the substrate as in the experimentally realized $7\times7$ supercell. We find that the electronic structure is affected by the substrate in similar ways for the two supercells. In both cases,  the occupied part of the Ta $d$ conduction band roughly follows the freestanding band with a small downward shift (circled in blue in Fig. \ref{fig:band-unfold}), but the top of the band flattens out well below the freestanding band for unoccupied states near and at $\Gamma$ (circled in red in Fig.  \ref{fig:band-unfold}) due to an avoided crossing with a Au band. In both cases, there are also prominent avoided crossings just above the Fermi level along $\Gamma$-$K$ and $\Gamma$-$M$, but these are artifacts of using a finite Au slab.\cite{supplemental} In Fig. \ref{fig:Au-PBands} we see that the conduction band is indeed primarily Ta $d$ in character, but exhibits non-negligible Au character, especially above the Fermi level near $\Gamma$ and below the Fermi level near M and K. The changes in the conduction band, including the band flattening near $\Gamma$ and the downward shift of the bottom of the band clearly arise from hybridization between monolayer and substrate states, and can not simply be modeled as a rigid shift. 
The actual substrate-monolayer charge transfer is
less than the doping deduced from the shift in occupied states, as discussed in Ref. \onlinecite{Wehling}, and is unlikely to contribute significantly to  the suppression of the CDW.  
\par

Interestingly, the Au substrate has a similar effect on the electronic structure whether we use the fully relaxed structure, the laterally constrained structure, or a rigid TaS$_2$ monolayer in proximity to the Au(111) surface. We also find that there are no major qualitative differences when van der Waals corrected functionals are employed, where the substrate-monolayer separation is larger. 
This indicates that the primary changes in the electronic structure are not caused by the lateral or vertical displacements of atoms in the TaS$_2$ monolayer, but rather by the reduction in symmetry at the interface and the resulting substrate-monolayer hybridization. This hybridization then drives
the structural distortion of the monolayer. \par

\subsection{TaS$_2$ on Graphene}

\begin{figure}[b]
  \includegraphics[width=0.85\linewidth,clip]{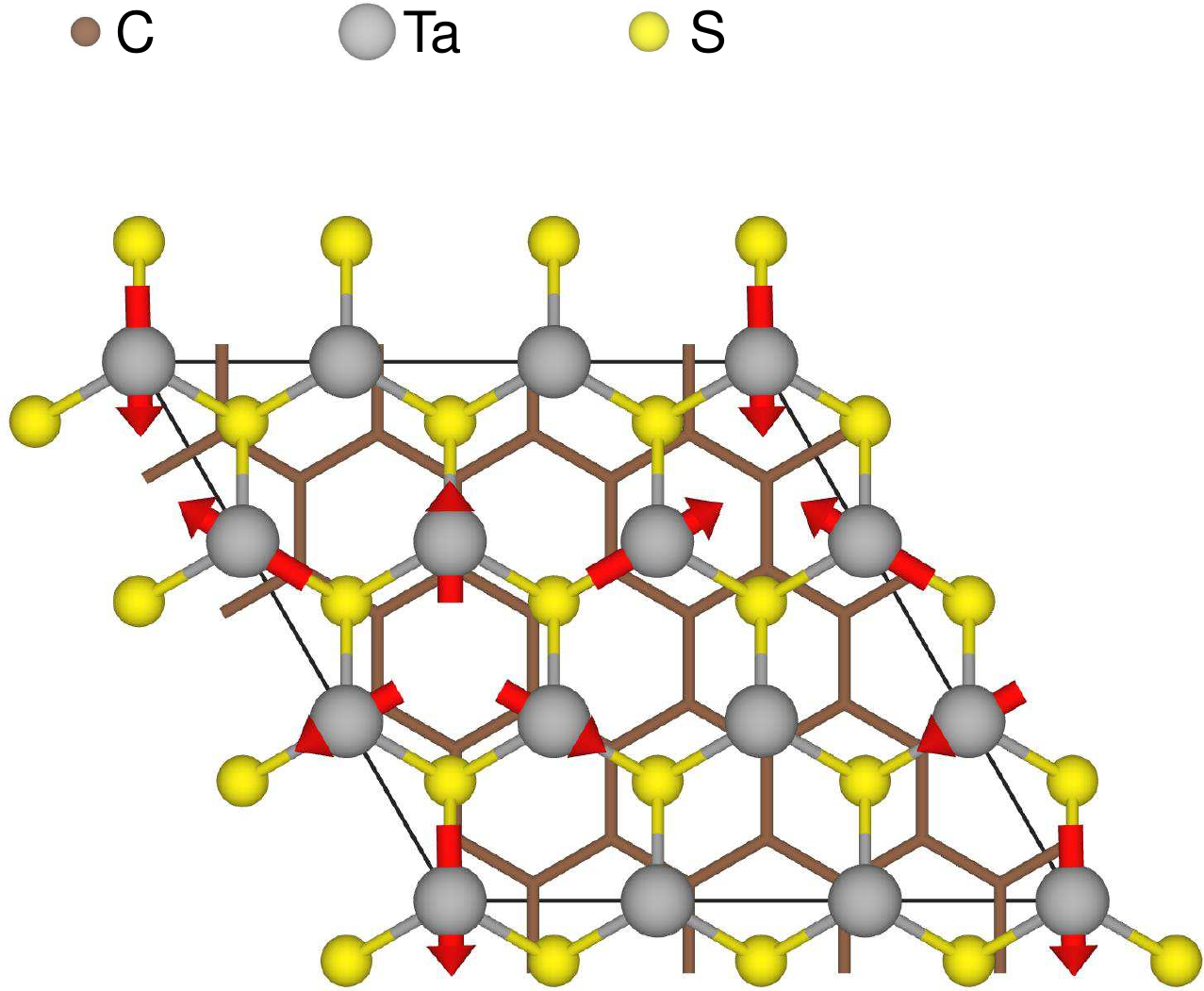}
  \caption{Top view of CDW distortion in $3\times 3$ 1H-TaS$_2$ monolayer on a graphene substrate. Arrows indicate lateral displacements of Ta atoms (not to scale).}
  \label{fig:Graphene-CDW}
\end{figure}

Since the electronic structure of the TaS$_2$ monolayer is altered significantly by the Au(111) substrate, it is perhaps not surprising that the CDW instability is affected by the substrate. For substrates that interact more weakly with the TaS$_2$ monolayer we would expect the CDW instability to remain, as observed experimentally on bilayer graphene.\cite{Lin-MBE-TMDonGraphene}

\begin{figure}[t]
  \includegraphics[width=0.85\linewidth,clip]{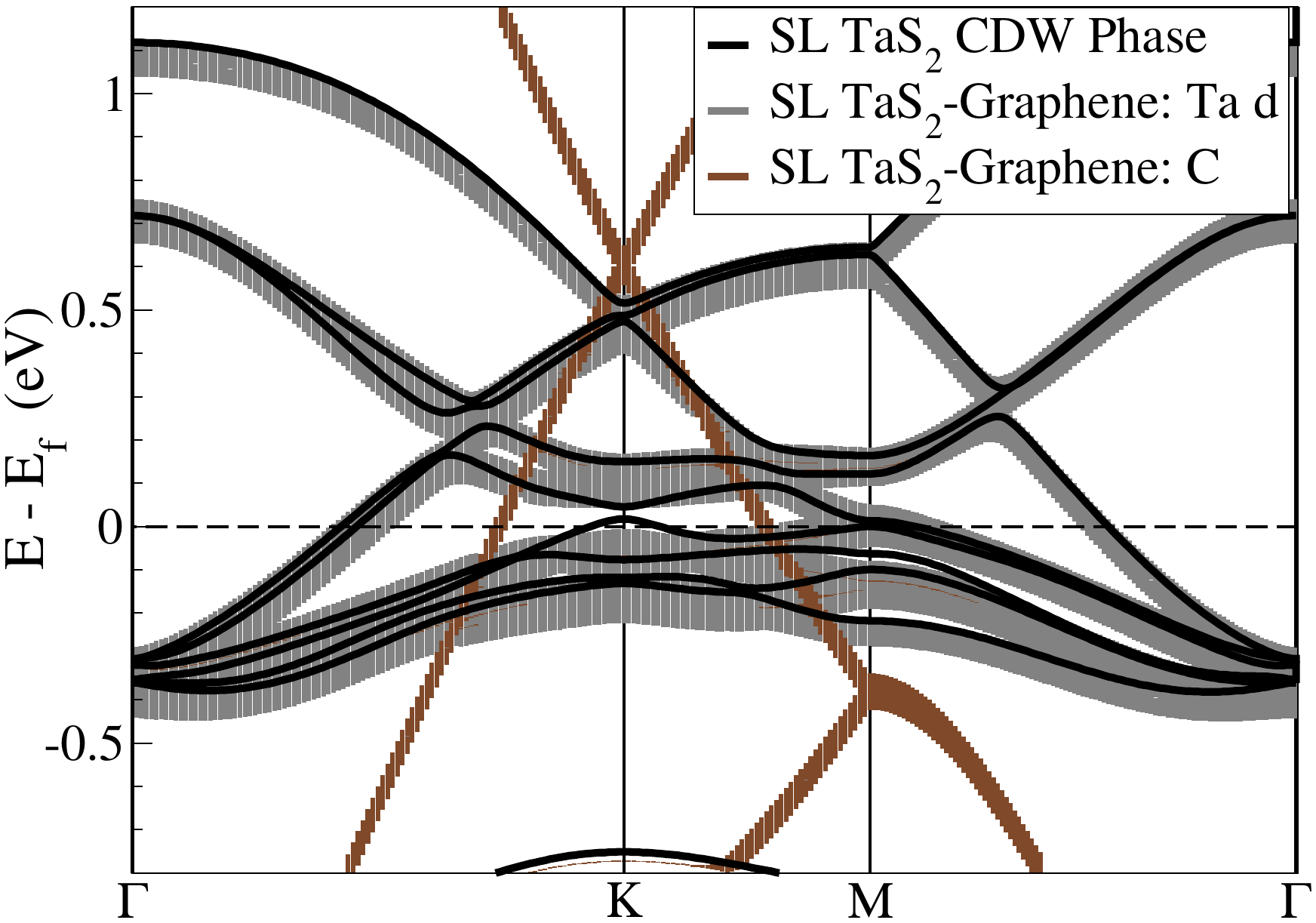}
  \caption{Orbital projected band structure of $3\times 3$ TaS$_2$ monolayer on graphene supercell. Bands of free-standing 1H-TaS$_2$ monolayer (with the CDW distortion expressed) are plotted with solid lines.}
  \label{fig:Graphene-PBands}
\end{figure}

Our calculations indeed find the $3 \times 3$ CDW phase of the 1H-TaS$_2$ monolayer to be robust on a graphene substrate, as shown in Fig. \ref{fig:Graphene-CDW}.
The Ta displacements are similar in both direction and magnitude to those  calculated for the CDW phase of the freestanding monolayer, with an average Ta displacement of 0.045 \AA. The displacement pattern is also similar to the CDW structure in 2H-TaSe$_2$ as determined from electron diffraction.\cite{Bird-2HTaSe2}
The calculated band structure for the 1H-TaS$_2$-graphene system is shown in Fig. \ref{fig:Graphene-PBands}. In contrast to the 1H-TaS$_2$-Au(111) system, there is minimal hybridization between the monolayer and substrate states. The structure of the Ta $d$ bands follows closely that of the freestanding monolayer in the CDW phase shown in Fig. \ref{fig:Graphene-CDW}, with a slight downward shift in energy. The bands retain their orbital character and do not develop avoided crossings. Based on the $\sim 0.6$ eV upward shift of the graphene Dirac point, we estimate that the TaS$_2$ monolayer becomes electron doped by about .036 electrons/Ta. This is well below the doping needed to suppress the CDW phase according to previous DFT studies.\cite{Albertini-TaS2doping} The graphene substrate has minimal impact on the electronic structure of the TaS$_2$ monolayer and preserves the CDW instability. \par

\section{\label{sec:conlusion}Conclusions}
Our calculations show that there is strong hybridization between the Au valence states and the S $p$ and Ta $d$ states of the 1H-TaS$_2$ monolayer. This interaction significantly alters the electronic structure of the monolayer, and this, as opposed to substrate-induced doping, is likely responsible for suppressing the CDW phase.
This  interaction also induces a significant structural distortion in the monolayer, unrelated to the CDW. The distortion pattern has the periodicity of the commensurate interface and is characterized by lateral and vertical displacements of interface S atoms that narrow the distribution of S-Au distances. Our rendered STM image and its FFT suggest that evidence for this distortion is present in experimental STM data. 
It is possible that similar distortions exist in other S-based transition metal
dichalcogenides grown epitaxially on Au. 

Our calculations also show that SL 1H-TaS$_2$ exhibits a robust CDW phase on graphene, and that its electronic structure undergoes minimal change. This is in accord with the experimental observation of a CDW phase in SL 1H-TaS$_2$ epitaxially grown on bilayer-graphene.\cite{Lin-MBE-TMDonGraphene} 
However, a recent experiment reports that SL 1H-TaS$_2$ exfoliated and encapsulated in hexagonal boron nitride does not exhibit a CDW phase.\cite{Kaxiras} The contrasting behavior observed on graphene and h-BN is surprising and warrants further investigation.

More generally, this work underscores the importance of separating substrate-induced effects and intrinsic properties of two-dimensional materials. 
While the choice of substrate may cause unintended changes to intrinsic properties of interest, it can offer opportunities to tune the properties of the system and induce new phases.

\begin{acknowledgments}
This work was supported by NSF Grant EFRI-1433307.
\end{acknowledgments}

%

\end{document}


\title{Supplemental Material: Substrate-induced suppression of charge density wave in monolayer 1H-TaS$_2$ on Au(111)}
\author{Heraclitos M. Lefcochilos-Fogelquist}
\affiliation{
    Department of Physics,
    Georgetown University,
    Washington, DC 20057, USA
}
\author{Oliver R. Albertini}
\affiliation{
    Department of Physics,
    Georgetown University,
    Washington, DC 20057, USA
}
\author{Amy Y. Liu}
\affiliation{
    Department of Physics,
    Georgetown University,
    Washington, DC 20057, USA
}
\maketitle

\begin{figure}[t]
\raggedright
\subcaptionbox{Orbital projected band structure of SL TaS$_2$-Au(111) using 5 Au layers}
{
\begin{tikzpicture}
\node(a){\includegraphics[width=0.46\textwidth]{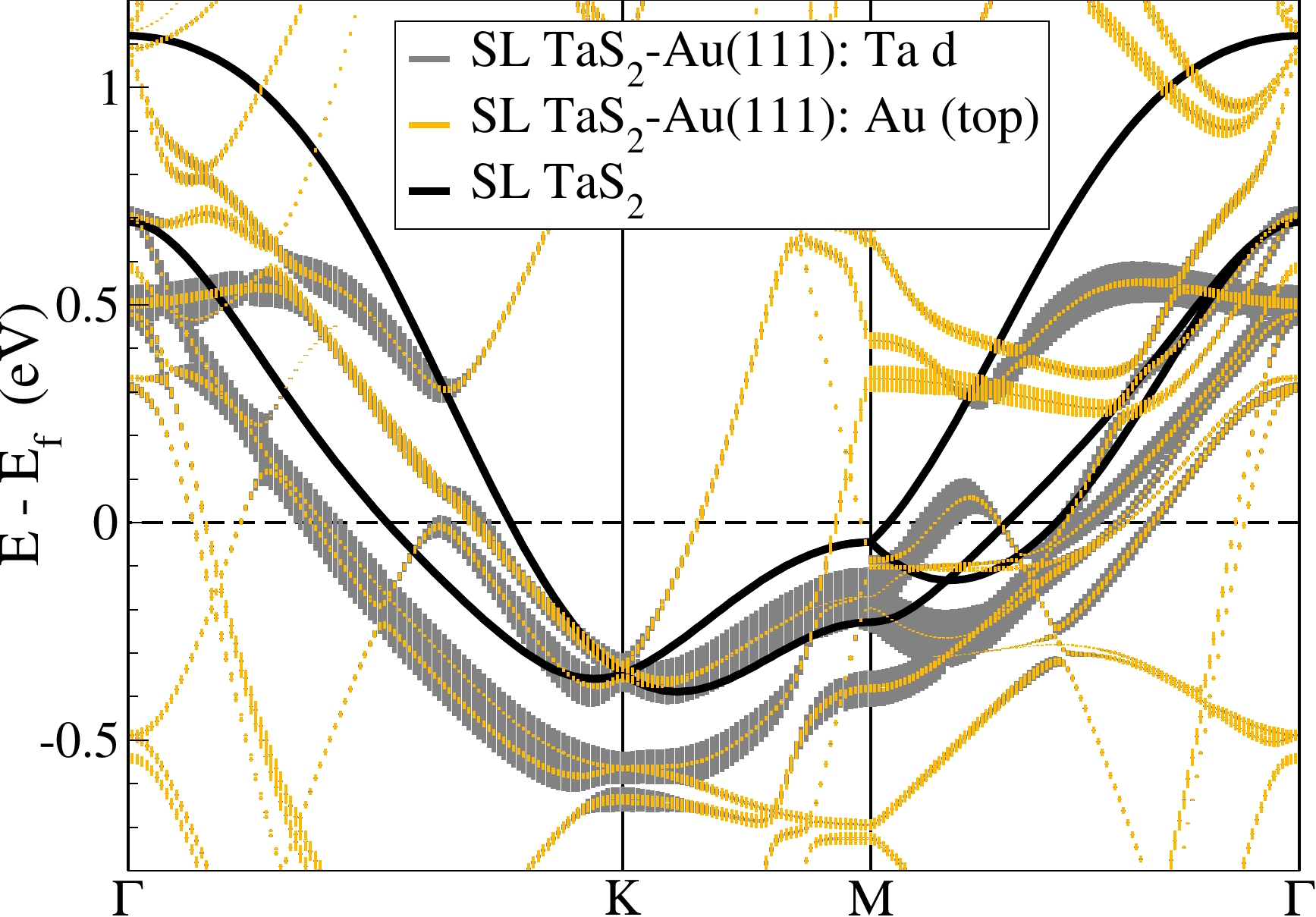}};
\node at(a.center)[draw, red, line width=1pt, ellipse, minimum width=20pt, minimum height=30pt, xshift=-37pt]{};
\node at(a.center)[draw, red, line width=1pt, ellipse, minimum width=20pt, minimum height=30pt, xshift=52pt]{};
\end{tikzpicture}  
}%
\hfill
\subcaptionbox{Orbital projected band structure of SL TaS$_2$-Au(111) using 32 Au layers} 
{
\begin{tikzpicture}
\node(a){\includegraphics[width=0.46\textwidth]{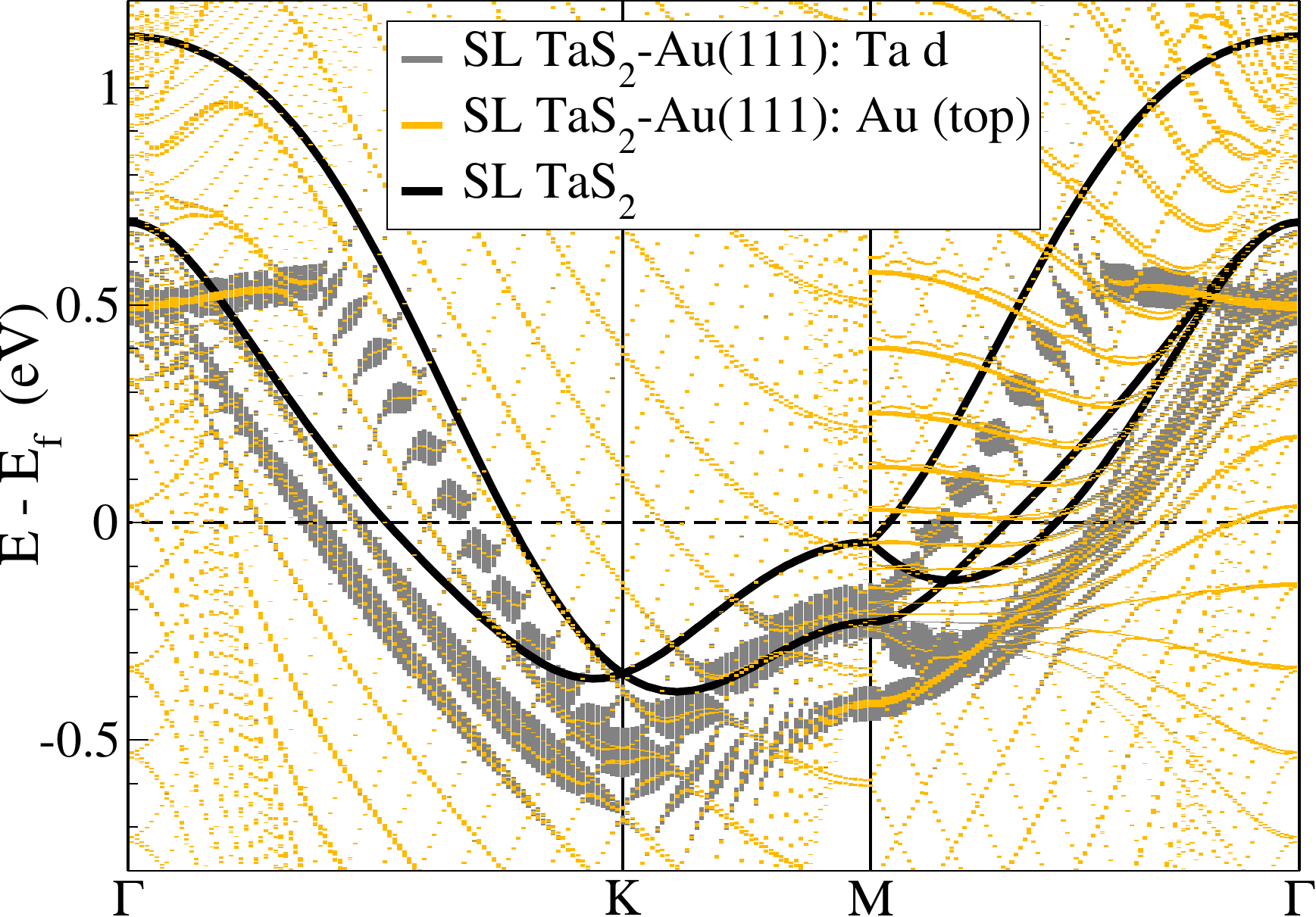}};
\end{tikzpicture}  
}%
\end{figure}

{\bf Figure 1:} Orbital projected band structure for TaS$_2$-Au(111) using (a)  a 5 layer Au slab  and (b) a 32 layer Au slab. Circled in red in (a) are the prominent spurious avoided crossings that give rise to the valley just above the Fermi level in the pDOS. Also visible are the quantized Au $k_z$ subbands which couple with the Ta $d$ conduction band leading to the spurious avoided crossings. In (b) we see that additional Au layers and a larger $c$ supercell lattice constant yield many more of these $k_z$ subbands, leading to the opening of more, but smaller, gaps in the Ta $d$ conduction band. This effect results in the progressive recovery of the freestanding band structure in certain regions, but not near and at $\Gamma$ above the Fermi level. This enables us to identify the band flattening and presence of Au character near $\Gamma$ as real effects.